\documentstyle[preprint,aps]{revtex}
\input epsf.tex
\def\DESepsf(#1 width #2){\epsfxsize=#2 \epsfbox{#1}}
\begin{document}

\preprint{\vbox{\hbox{DOE-ER40757-097
   }\hbox{UTEXAS-HEP-97-6}\hbox{OITS-624}\hbox{OSURN-323}}}
\draft
\title {Scalar tau Signal at LEP 2 in models with Gauge Mediated Supersymmetry
Breaking}
\author{ D. A. Dicus$^{1,2} $, B. Dutta$^{3} $, and S. Nandi$^{4}$ }
\address{$^{1} $ Center for Particle Physics,
 University of Texas, Austin, TX 78712\\$^{2} $ Department of Physics, University
of Texas, Austin, TX 78712\\$^{3} $ Institute of Theoretical Science, University
of Oregon, Eugene, OR 97403\\$^{4} $ Department of Physics, Oklahoma State
University, Stillwater, OK 74078}
\date{April, 1997)\\(To appear in Phys. Rev. D}
\maketitle
\begin{abstract} In theories with gauge mediated supersymmetry breaking, the
scalar tau, (${\tilde
\tau_1}$) is the lightest observable superpartner for part of the parameter
space. At LEP 2, the production of such a
$\tilde \tau_1$ pair and their subsequent decays give rise to a pair of $\tau$
leptons plus missing energy from the unobserved gravitinos. The angular
distributions of the $\tau$'s are different from those arising from the
production and decay of W pairs, and thus will constitute an interesting signal
for supersymmetry. We also consider ${\tilde \tau_1}$ pair production in the
complementary part of parameter space where the lightest neutralino is lighter
than the
$\tilde \tau_1$.
\end{abstract}
\pacs{PACS numbers: 11.30.Pb  12.60.Jv 14.80.Ly}
\newpage In spite of many interesting features of supersymmetric theories, the
mechanism of supersymmetry breaking and how it is communicated to the observable
sector has been a major area of concern for over a decade. In most of the
previous work it has been assumed that the supersymmetry is broken in a hidden
sector at a scale of
$\sim 10^{11}$ GeV, and communicated to the observable sector via the
gravitational interaction. However, this scenario has a major problem involving
flavor changing neutral current (FCNC). This can be avoided if supersymmetry is
broken in a hidden sector at a scale
$10^{5}$ GeV, and communicated to the observable sector by means of the Standard
Model (SM) gauge interaction\cite{DN}. The number of parameters is also smaller
in these gauge mediated models (GMSB) which makes them more predictive. These
models have a very distinctive feature, the gravitino is the lightest
supersymmetric particle(LSP) and all super particles must ultimately decay to it.
This feature gives rise to some new signals in colliders that may be observed in
the near future.

  One interesting aspect of gauge mediated models is that the role of the
next-to-lightest supersymmetric particle (NLSP) can be taken by either the
lighter scalar tau ($\tilde \tau_1$) or the lightest neutralino.  In the past
year a lot of work has been done
\cite{{DWR},{swy},{bkw},{akkmm},{dwt},{bbct},{BPM},{rtm},{BR},{aab}}in these
models motivated in part by the fact that, when the neutralino is the NLSP, the
CDF event (ee$\gamma\gamma$+missing energy \cite{SP}) can be explained. However
there is a vast region of parameter space where  the lighter stau  can be the
NLSP and the lightest neutralino is the NNLSP(Next to Next to LSP). Recently, we
considered neutralino pair production in these scenarios  where the stau is the
NLSP and showed that the neutralino pair gives rise to 4$\tau$ + missing energy
in the final state without any standard model background \cite{we}. In this paper
we look at direct production of stau in the scenarios where the lighter stau is
the NLSP as well as in the scenarios where neutralino is the NLSP. In the first
case, since the only decay mode available to a lighter stau is to decay into a
$\tau$ and a gravitino, the signal is 2
$\tau$ + missing energy.  We compare the signal with the background generated
from W pair production (each W can decay into a $\tau$ and a tau
neutrino($\nu_{\tau}$)), and show that the angular distribution of the individual
$\tau$'s will make the signal look  very different from the background. For the
2nd case the signal is 2 $\tau$ +2$\gamma$ with missing energy since each stau
decays into a neutralino and a
$\tau$ and the neutralino then decays into a hard photon and a gravitino. So the
signal itself can discern the two cases. We also discuss the angular distribution
and energy distribution of the decay products in the 2nd case.

In  GMSB models the superparticle masses depend on five parameters
$M,\Lambda,n,\tan\beta, \mu$. $M$
 is the messenger scale, $M=\lambda <s>$, where $<s>$ is the VEV of the scalar
component of the hidden sector superfields, and $\lambda$ is the Yukawa coupling.
The parameter $\Lambda$ is equal to $<F_s>/<s>$, where
$<F_s>$ is the VEV of the auxillary component of $s$. $F_s$ can be $\sim F$
\cite{IT} , where $F$ is the intrinsic SUSY breaking scale. In GMSB models,
$\Lambda$ is taken around 100 TeV, so that the colored superpartners have masses
around a TeV or less. The parameter n is fixed by the choice for the messenger
sector. The messenger sector representations should be vector like (for example,
$5+{\bar 5}$ of
$SU(5)$, $10+{\bar {10}}$ of
$SU(5)$ or $16+{\bar{16}}$ of $SO(10)$) so that their masses are well above the
electroweak scale. They are also chosen to transform as a GUT multiplet in order
not to affect the gauge coupling unification in MSSM. This restricts $n(5+{\bar
5})\le 4$, $n(10+{\bar {10}})\le 1$ in
$SU(5)$,  and $n(16+{\bar {16}})\le 1$ in
$SO(10)$ GUT for the messenger sector (one
${10}+{\bar{10}}$ pair corresponds to
$n(5+{\bar 5})$=3). The parameter $\tan\beta$ is the usual ratio of the up
($H_u$) and down ($H_d$) type Higgs VEVs. The parameter
$\mu $ is the coefficient in the bilinear term, $\mu H_uH_d$ in the
superpotential, while another parameter $B$ is defined to be the coefficient in
the bilinear term,
$B\mu H_uH_d$ in the potential. In general, $\mu$ and $B$ depend on the details
of the SUSY breaking in the hidden sector. We demand that the electroweak
symmtery is broken radiatively which determines $\mu^2$ and $B$ in terms of the
other parameters of the theory. Thus we are left with five independent parameters,
$M,\Lambda,n,\tan\beta$ and sign($\mu$). The soft SUSY breaking gaugino and the
scalar masses at the messenger scale M are given by \cite{{DN},{SPM}}
\begin{equation}
\tilde M_i(M) = n\,g\left({\Lambda\over M}\right)\,
{\alpha_i(M)\over4\pi}\,\Lambda.
\end{equation} and
\begin{equation}
\tilde m^2(M) = 2 \,(n)\, f\left({\Lambda\over M}\right)\,
\sum_{i=1}^3\, k_i \, C_i\,
\biggl({\alpha_i(M)\over4\pi}\biggr)^2\,
\Lambda^2.
\end{equation} where $\alpha_i$, $i=1,2,3$ are the three SM gauge couplings and
$k_i=1,1,3/5$ for SU(3), SU(2), and U(1), respectively. The $C_i$ are zero for
gauge singlets, and 4/3, 3/4, and $(Y/2)^2$ for the fundamental representations of
$SU(3)$ and $SU(2)$ and
$U(1)_Y$ respectively (with Y defined by $Q=I_3+Y/2)$. Here $n$ corresponds to
$n(5+{\bar 5})$. $g(x)$ and
$f(x)$ are messenger scale threshold functions with $x=\Lambda/M$.

We have calculated the SUSY mass spectrum using the appropriate RGE equations
\cite{BBO} with the boundary conditions given by equation (1) and (2), and
varying the five free parameters. Although in principle the messenger scale is
arbitrary (with
$M/\Lambda>1$), in our analysis we have restricted $1<M/\Lambda<10^4$ and chosen
$\Lambda\sim 100$ TeV.
 For the messenger sector, we choose  $5+{\bar 5}$ of SU(5), and varied
$n(5+{\bar 5})$ from 1 to 4. In addition to the current experimental bounds on
the superpartner masses, the rate for $b\rightarrow s\gamma $ decay puts useful
constraints on the parameter space \cite{{dwt},{bbct},{ddo}}. It is  found that
\cite{dwt,BPM} for $n=1$ and low values $\tan\beta$ ($\tan\beta\le25$), the
lightest neutralino
$\chi_0$ is the NLSP for $M/\Lambda>1$. As $\tan\beta$ increases,
$\tilde\tau_1$ becomes the NLSP for most of the parameter space with lower values
of
$\Lambda$. For $n\ge2$, $\tilde\tau_1$ is the NLSP even for the low values of
$\tan\beta$ (for example, $\tan\beta\gtrsim 2$), and for $n\ge3, $
$\tilde\tau_1 $ is again naturally the NLSP for most of the parameter space. In
Tables 1 and 2 we show five sets of spectrum where the lighter stau is the NLSP
and five sets where the lightest neutralino is the NLSP. We use these scenarios
for detailed calculations. 

Let us first discuss the production cross section of the $\tilde\tau_1$ pair in
the scenarios 1-5 of Table 1, where the lighter stau is the NLSP. The total
cross-sections 
 are given in Table 3 for three LEP2 energies, $\sqrt s$ =172, 182 and 194 GeV .
Each of the produced
$\tilde\tau_1$ will decay into a
$\tau$ and a gravitino with essentially a 100$\%$ branching ratio.  Thus, from
$\tilde\tau_1$ pair production, we obtain  final states with two $\tau$ 's and
missing energy. For example, the number of events in scenario 5 for $\sqrt s$
=182 GeV, with 100$pb^{-1}$ luminosity, is 27, and for $\sqrt s$ =194 GeV , with
 250$pb^{-1}$ luminosity is 70. The decay,
$\tilde\tau_1\rightarrow\tau {\tilde G}$  is fast enough so that it takes place
inside the detector. (If
$\sqrt F$ is much larger than a few 1000 TeV \cite{new} , then $\tilde\tau_1$
will decay outside the detector. In that case  the signal will be 2 heavy charged
particles passing through the detector).

There is a considerable background from W production, where each W decays into a
$\tau$ and a neutrino. However since the staus are scalar particles, their
angular distribution is different from the W distribution. This shows up in the
individual $\tau$ angular distributions. In Figure 1 we show these angular
distributions of the
$\tau$s. As expected the angular distribution of the  $\tau$ coming from stau
decay is basically flat; the same distribution from W decay has significant
$\theta$ dependence (as measured from the beam direction). Consequently, the
signal is almost twice the background for $\tau^{-}$ when cos$\theta$ is
negative. The opposite thing happens for
$\tau^{+}$ i.e the positive values of the cos$\theta$ regions need to be searched
for $\tau^{+}$ in order to extract the signal.

	The information in Fig.1, and its mirror image for the $\tau$ of the other
charge, is made more precise in Table 4 where we give the fraction of events that
should be found in each bin of width 0.2 in cos$\theta_+$ and in cos$\theta_-$.

	Figure 2 shows the distribution in the angle between the two tau is very similar
in stau or W production.

	In Table 3 we also show the cross-section for charged Higgs production
 in a usual SUSY model. The charged Higgs  in the GMSB model are too heavy to be
produced at LEP 2 but in other models they could be light and, because they decay
into a
$\tau$ + $\nu_{\tau}$ with essentially 100$\%$ branching ratio if tan$\beta$ is
greater than 2, and because they are produced with the same angular distribution
as the stau, they are indistinguishable from stau production of the same mass in
GMSB except through the total production rate which is a little larger than for
stau. For example the number of events, when the charged Higgs mass is comparable
to the lighter stau mass in the scenario 5, is  34 for
$\sqrt s$ =182 GeV with 100$pb^{-1}$ luminosity and is 89 for
$\sqrt s$ =194 GeV with 250$pb^{-1}$ luminosity. So if a two tau signal is
observed at LEP 2, with a limited number of events so that the difference in the
production rate is not sufficient to discern between these two models, then other
signals, such as 4 $\tau$ + missing energy from neutralino pair production
\cite{we}, will have to be used. We note here that in the case of charged Higgs
pair productions, with subsequent decay into $\tau$ will have similar angular
distribution as those coming from stau. Thus, the signal for the charged Higgs
can also be distinguished from the W-background using the distribution shown in
Fig1. 

We now discuss the scenarios of Table 2 where the lightest neutralino is the
NLSP. In Table 5 we give the cross section where staus are produced. The signal
in this case is different than the signal in the case where stau is the NLSP;
here each stau will decay into a
$\tau$ and a neutralino and the neutralino will then decay into a gravitino and 
a photon giving rise to 2 $\tau$ + 2 $\gamma$ + missing energy in the final
state. The photons are hard and they can be easily seperated from 
bremsstrahlung. The number of events in scenario 6 for
$\sqrt s$ =182 GeV, with 100$pb^{-1}$ luminosity, is 30 and  for $\sqrt s$ =194
GeV , with 250$pb^{-1}$ luminosity is 77. Since the neutralino is the NLSP in
these scenarios, they can be pair produced more easily than the staus.
Consequently the signal of  2 photons(hard) and missing energy in the final state
coming from the neutralino decays could be used to discover SUSY and a detailed
analysis of the stau production will then give  information on particular models.

 We now give three graphs showing the angles and energies of the final state
$\tau$s and photons. There is nothing very surprising in these plots but we
include these because they might be useful in verifying that detected $\tau$s and
photons are from this process.In Figure 3 we plot the angular distributions of
the one of the emitted photons and the distribution of the angle between the two
emitted photons. The angular distribution of the single photon is almost
isotropic while the angular distribution of the angle between the photons has
$\theta$ dependence; the distribution is large when cos$\theta$ is negative. The
angular distribution of one of the $\tau$, or of the angle between the two
$\tau$ looks very similar to Fig. 3.  In Fig. 4 we show the angle between a tau
and a photon from the same stau decay and between a tau and a photon which come
from opposite stau decays. In Fig. 5, we also plot the energy distributions of
the decay products, ie., the total missing energy, the energy of one of the
photons and the energy of one of the $\tau$'s. Since the massive neutralino
carries more than half of the energy of the lighter stau, each gravitino carries
more than 1/6 of the total energy and the total missing energy distribution
becomes a maximum at more than 1/3 of the beam energy. The energy distribution
for the
$\tau$'s and photons maximize at much smaller energy.

In this paper we concentrated on the production of staus. There are other
scalar particles, selectron and smuon, whose lighter component's masses can
be less than the beam energy but this is usually not the case for LEP II.
Even when they are lighter their masses are greater than that of the stau and
close to the beam energy so that their production is suppressed relative to
that of the stau. Production of selectron is further suppressed by the
additional contribution of neutralino exchange in the t-channel while the
production rate of the smuons is the same except for the effects of the
different (larger) mass. There are three possible signals for the production
of selectrons (smuons): 1) $E_b>m_{\tilde e}>m_{\chi^0}>m_{\tilde \tau}$,
where $E_b$ is the beam energy, then the $\tilde e$ will decay to $e+\chi^0$
followed by $\chi^0\rightarrow \tau+\tilde\tau$ and $\tilde \tau\rightarrow
\tau \tilde G$. The final state is $e^+ +e^- + 4\tau$ plus missing energy. 2)
If $E_b>m_{\tilde e}$ and $m_{\chi^0}>m_{\tilde e}$ then the selectron decays
directly to an electron and a gravitino; the final state is just $e^+e^-$
plus missing energy. 3) If $E_b>m_{\tilde e}>m_{\tilde
\tau}>m_{\chi^0}$ then the selectron will decay to $e+\chi^0$ with
$\chi^0->\gamma +\tilde G$; the final state is $e^+e^-+2\gamma$ plus missing
energy.

In conclusion, we have discussed the production of scalar staus and their decay
modes in gauge mediated models in the scenarios where the lighter stau is the
NLSP as well as in the scenarios where the neutralino is the NLSP. We have 
discussed the background for the individual cases and have shown how the signal
can be extracted from the SM background. We also have noted that the charged
Higgs in the supergravity models can have signatures that are almost identical to
that of the stau in these models so that observation of this signal is a sign of
new physics but not necessarily of gauge mediated supersymmetry breaking.
 
After finishing this work, we came across a preprint by   S. Ambrosanio,  G. D.
Kribs and  S. P. Martin, hep-ph/9703211, which has partial overlap with our work.

We are very grateful to David Strom of OPAL collaboration for many discussions.
Part of this work has been done when one of us (S.N.) was on sabbatical leave at
the University of Texas at Austin. He wishes to thank Duane Dicus of the Center
for Particle Physics for very warm hospitality and support during his stay there.
This work was supported in part by the  US Department of Energy Grants No.
DE-FG013-93ER40757, DE-FG02-94ER40852, and DE-FG03-96ER-40969.

\newpage
\begin{center} {\bf TABLE CAPTIONS}\end{center}
\begin{itemize}
\item[Table 1~:] {Mass spectrum for the superpartners in some scenarios where the
lighter stau is the NLSP.  Note that the 1st and 2nd generation superpartner
masses are almost same}.
\item[Table 2~:] {Mass spectrum for the superpartners in some scenarios where the
lightest neutralino is the NLSP}.
\item[Table 3~:] { For each scenario in Table 1 and three  beam energies the
first line gives the total cross section for stau pair production, while the 2nd 
line gives the total cross section for charged Higgs production.}
\item[Table 4~:] { The double angular distribution of the two tau. The numbers
are the per cent of total events where each tau is in a bin in cos$\theta$ of
width 0.2.  The first number in each bin is for $\tau$ from
$e^{+}e^{-} \rightarrow\tilde \tau_1\tilde \tau_1\rightarrow
\tau\tau\tilde G\tilde
 G$; the number in brackets is for $\tau$ from $e^{+}e^{-}
\rightarrow WW\rightarrow\tau\tau
\nu_{\tau}\nu_{\tau}$}. The table is symmetric about the complementary diagonal,
ie., the numbers for (cos$\theta_{+}$ , cos$\theta_{-}$) are the same as those
for (- cos$\theta_{-}$ , - cos$\theta_{+}$).
\item[Table 5~:] { For each scenario in Table 2 and three beam energies the total
cross section for stau pair production is shown.}
\end{itemize}
\begin{center} {\bf FIGURE CAPTIONS}\end{center}
\begin{itemize}
\item[Fig. 1~:] {The angular distribution of $\tau^{-}$, relative to the electron
beam axis, from the stau decay (dashed line) and from W decay (solid line). The
distribution of the
$\tau^{+}$ is the mirror image of this graph.}
\item[Fig. 2~:] {The relative number of events as a function of the angle between
the $\tau^{+}$ and the $\tau^{-}$. The dashed line corresponds to the
$\tau$s from the stau decay and the solid line corresponds to the $\tau$s from
the W decay.}
\item[Fig. 3~:] {The angular distribution of one of the photons (dashed line)
 and the angle between the two photons (solid line) from neutralino decay, when
the neutralino is the NLSP.}
\item[Fig. 4~:] {The angular distribution between a $\tau$ and a photon.
 The solid line corresponds to the $\tau$ and the photon from the same stau and
the dashed line corresponds to the same particles from opposite stau.}
\item[Fig. 5~:] {The distribution of the missing energy (solid line), the energy
of one of the photons (dashed), and the energy of one of the
$\tau$ (dot-dashed), when the neutralino is the NLSP.}

\end{itemize}

\newpage
\begin{center}  Table 1 \end{center}
\begin{center}
\begin{tabular}{|c|c|c|c|c|c|}  \hline &Scenario 1&Scenario 2&Scenario 3&Scenario
4&Scenario 5\\\hline masses&$\Lambda=63.7$ TeV,&$\Lambda=33$ TeV,&$\Lambda=60$
TeV,&$\Lambda=59.7$ TeV,&$\Lambda=28$ TeV,\\ (GeV)&n=1,
$M=4\Lambda$&n=2, $M=20\Lambda$&n=1, $M=10\Lambda$&n=1,
$M=10\Lambda$&n=2,
$M=40\Lambda$\\
&$\tan\beta$=31.5&$\tan\beta$=20&$\tan\beta$=31.5&$\tan\beta$=28.5&
$\tan\beta$=18
\\\hline m$_h$&121&117&120&120&114\\\hline
m$_{H^{\pm}}$&366&318&356&364&278\\\hline m$_A$&357&308&347&355&266\\\hline
m$_{\chi^0}$&85&87&80&80&72\\\hline m$_{\chi^1}$&158&156&149&148&128\\\hline
m$_{\chi^2}$&350&286&345&343&249\\\hline m$_{\chi^3}$&364&309&358&356&275\\\hline
m$_{\chi^{\pm}}$&157,367&155,312&149,361&127,277&158,367\\\hline m$_{\tilde
{\tau}_{1,2}}$&74,249&73,192&65,240&74,236&65,167\\\hline m$_{\tilde
{e}_{1,2}}$&120,236&96,184&116,225&115,224&85,159\\\hline m$_{\tilde {\rm
t}_{1,2}}$&664,727&515,588&607,673&605,672&432,505\\\hline m$_{\tilde {\rm
b}_{1,2}}$&698,740&558,586&641,686&643,683&472,497\\\hline m$_{\tilde {\rm
u}_{1,2}}$&737,765&580,601&683,710&679,709&490,508\\\hline m$_{\tilde {\rm
d}_{1,2}}$&735,769&580,606&681,715&678,711&490,514\\\hline m$_{\tilde
g}$&565&587&533&530&498\\\hline
$\mu$&-343&-278&-337&-336&-240\\\hline
\end{tabular}
\end{center}

\newpage
\begin{center}  Table 2 \end{center}
\begin{center}
\begin{tabular}{|c|c|c|c|c|c|}  \hline &Scenario 6&Scenario 7&Scenario 8&Scenario
9&Scenario 10\\\hline masses&$\Lambda=42$ TeV,&$\Lambda=44$ TeV,&$\Lambda=41$
TeV,&$\Lambda=38$ TeV,&$\Lambda=42$ TeV,\\ (GeV)&n=1,
$M=60\Lambda$&n=1, $M=80\Lambda$&n=1, $M=1000\Lambda$&n=1,
$M=800\Lambda$&n=1,
$M=1200\Lambda$\\
&$\tan\beta$=21&$\tan\beta$=18&$\tan\beta$=15&$\tan\beta$=12&$\tan\beta$=8
\\\hline m$_h$&112&113&111&109&109\\\hline
m$_{H^{\pm}}$&280&300&296&277&316\\\hline m$_A$&268&290&285&265&306\\\hline
m$_{\chi^0}$&54&57&53&48&53\\\hline m$_{\chi^1}$&98&104&96&87&98\\\hline
m$_{\chi^2}$&255&270&263&244&274\\\hline m$_{\chi^3}$&272&286&281&263&293\\\hline
m$_{\chi^{\pm}}$&98,275&290,103&284,96&266,86&96,295\\\hline m$_{\tilde
{\tau}_{1,2}}$&62,174&73,180&78,171&77,157&90,171\\\hline m$_{\tilde
{e}_{1,2}}$&89,164&92,172&92,165&86,154&94,170\\\hline m$_{\tilde {\rm
t}_{1,2}}$&419,497&432,515&377,472&446,355&380,483\\\hline m$_{\tilde {\rm
b}_{1,2}}$&442,473&484,512&416,442&391,412&448,432\\\hline m$_{\tilde {\rm
u}_{1,2}}$&466,485&485,506&436,457&406,425&445,467\\\hline m$_{\tilde {\rm
d}_{1,2}}$&465,492&464,492&436,464&406,432&445,474\\\hline m$_{\tilde
g}$&374&392&366&339&375\\\hline
$\mu$&-244&-259&-253&-233&-375\\\hline
\end{tabular}
\end{center}

\newpage
\begin{center}  Table 3 \end{center}
\begin{center}
\begin{tabular}{|c|c|c|c|}  \hline Scenario 1&$\sqrt s$= & & \\ &
172(GeV)&182&194\\\hline 1& 0.117 pb&0.155&0.186\\ & (0.152)
pb&(0.198)&(0.235)\\\hline 2& 0.132&0.169&0.198\\ & (0.169)
&(0.214)&(0.248)\\\hline 3& 0.246&0.267&0.279\\ & (0.322)
&(0.344)&(0.355)\\\hline 4& 0.117 &0.155&0.186\\ & (0.152)
&(0.198)&(0.235)\\\hline 5& 0.250 pb&0.270&0.283\\ & (0.322)
pb&(0.344)&(0.356)\\\hline eeWW&0.177&0.212&0.225\\\hline
\end{tabular}
\end{center}
\newpage
\begin{center}  Table 4 \end{center}
\begin{center}
\begin{tabular}{|c|c|c|c|c|c|c|c|c|c|c|}  \hline
&\multicolumn{10}{c|}{cos$\theta_{-}$}\\\cline{2-11}
{cos$\theta_{+}$}&-0.9&-0.7&-0.5&-0.3&-0.1&0.1 &0.3&0.5&0.7&0.9\\\hline
-0.9&0.42&0.51&0.62&0.71&0.80&0.91&0.98&1.04&1.09&1.06\\
&(0.37)&(0.57)&(0.88)&(1.28)&(1.73)&(2.32)&(2.94)&(3.45)&(3.58)&(2.67)\\\hline
-0.7&0.51&0.63&0.75&0.86&0.96&1.08&1.16&1.19&1.18&\\
&(0.30)&(0.45)&(0.69)&(1.00)&(1.40)&(1.92)&(2.47)&(3.08)&(3.51)&\\\hline
-0.5&0.62&0.75&0.88&1.02&1.11&1.20&1.23&1.27&&\\
&(0.26)&(0.38)&(0.56)&(0.82)&(1.16)&(1.57)&(2.05)&(2.55)&&\\\hline
-0.3&0.71&0.86&1.02&1.13&1.23&1.29&1.30&&&\\
&(0.21)&(0.32)&(0.47)&(0.68)&(0.92)&(1.26)&(1.61)&&&\\\hline
-0.1&0.80&0.96&1.11&1.23&1.30&1.31&&&&\\
&(0.18)&(0.27)&(0.42)&(0.55)&(0.76)&(0.99)&&&&\\\hline
0.1&0.91&1.08&1.20&1.29&1.31&&&&&\\
&(0.16)&(0.23)&(0.38)&(0.45)&(0.58)&&&&&\\\hline 0.3&0.98&1.16&1.28&1.30&&&&&&\\
&(0.15)&(0.21)&(0.29)&(0.35)&&&&&&\\\hline 0.5&1.04&1.19&1.27&&&&&&&\\
&(0.15)&(0.19)&(0.23)&&&&&&&\\\hline 0.7&1.09&1.18&&&&&&&&\\
&(0.15)&(0.17)&&&&&&&&\\\hline 0.9&1.06&&&&&&&&&\\ &(0.15)&&&&&&&&&\\\hline
\end{tabular}
\end{center}
\newpage
\begin{center}  Table 5 \end{center}
\begin{center}
\begin{tabular}{|c|c|c|c|}  \hline Scenarios&$\sqrt s$= & & \\ &
172(GeV)&182&194\\\hline 6& 0.293 pb&0.307&0.312\\\hline 7&
0.132&0.169&0.198\\\hline 8& 7.22$\times10^{-2}$&0.114&0.151\\\hline 9&
8.54$\times10^{-2}$&0.127&0.162\\\hline 10&
-&4.39$\times10^{-3}$&4.04$\times10^{-2}$\\\hline
\end{tabular}
\end{center}

\newpage
\begin{figure}[htb]
\vspace{1 cm}

\centerline{ \DESepsf(fig12.epsf width 12 cm) }

\centerline{ \DESepsf(fig34.epsf width 12 cm) }

\centerline{ \DESepsf(fig5.epsf width 12 cm) }

\end{figure}
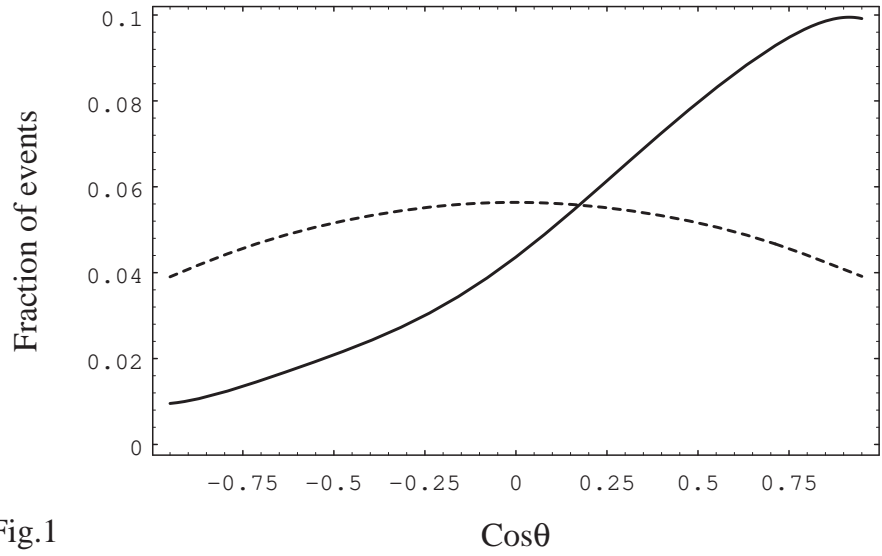
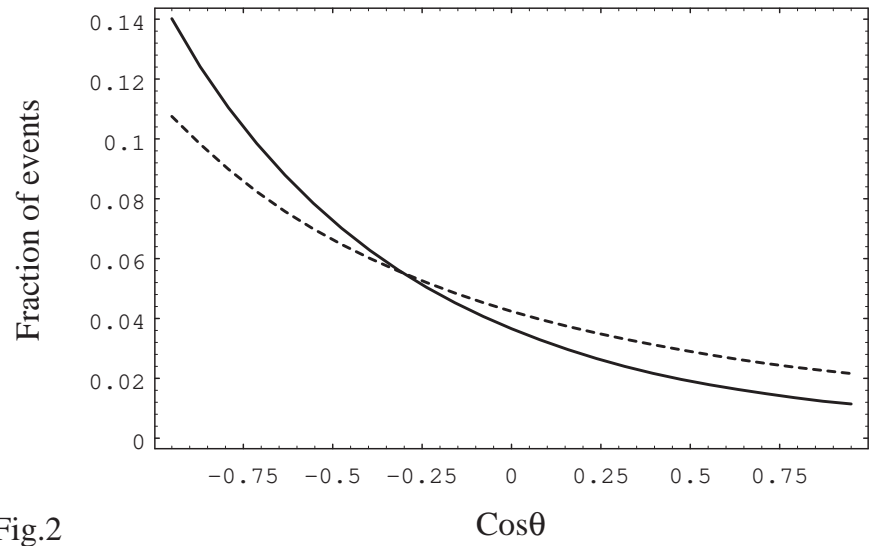
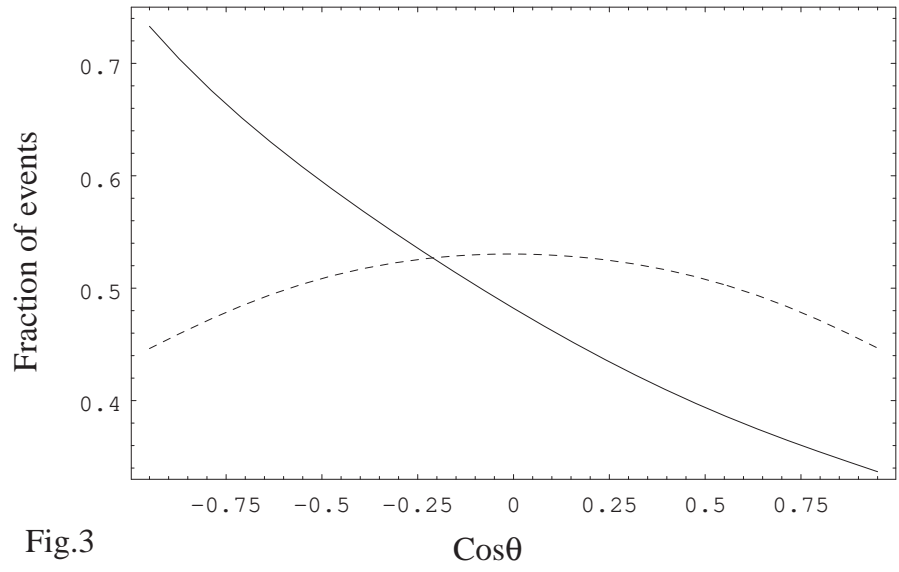
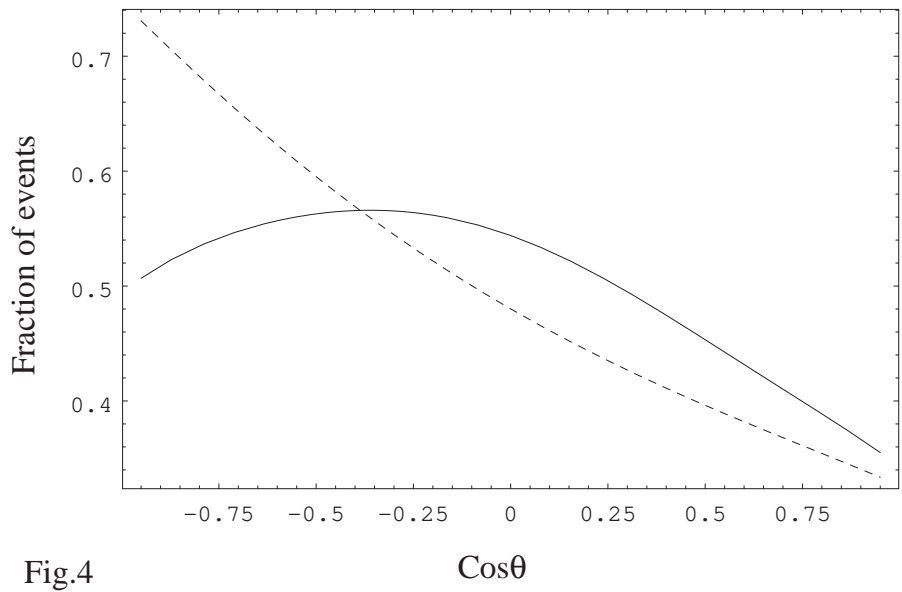
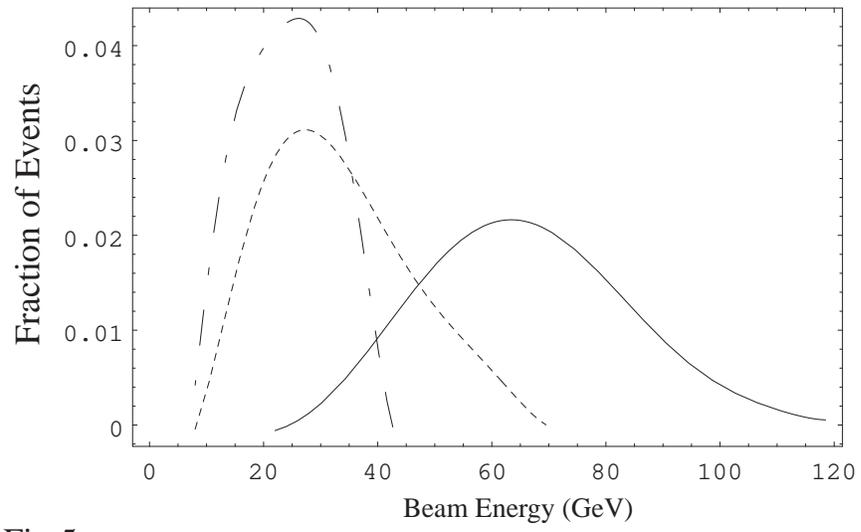
\end{document}